# Artificial Intelligence Enabled Spectral Reconfigurable Fiber Laser

YANLI ZHANG[†], SHANSHAN WANG[†], MINGZHU SHE, WEILI ZHANG[*]

*Fiber Optics Research Centre, School of Information and Communication Engineering, University of Electronic Science & Technology of China, Chengdu, 611731, China*
[†]*The first two authors contribute equally to this paper*
[*]*Corresponding author email: wl_zhang@uestc.edu.cn*

**Abstract:** The combinations of artificial intelligence and lasers provide powerful ways to form smart light sources with ground-breaking functions. Here, a Raman fiber laser (RFL) with reconfigurable and programmable spectra in an ultra-wide bandwidth is developed based on spectral-spatial manipulation of light in multimode fiber (MMF). The proposed fiber laser uses nonlinear gain from cascaded stimulated Raman scattering, random distributed feedback from Rayleigh scattering, and point feedback from an MMF-based smart spectral filter. Through wavefront shaping controlled by a genetic algorithm, light of selective wavelength(s) can be focused in the MMF, forming the filter that, together with the active part of the laser, actively shape the output spectrum with a high degree of freedom. We achieved arbitrary spectral shaping of the cascaded RFL (e.g., continuously tunable single-wavelength and multi-wavelength laser with customizable linewidth, mode separation, and power distribution) from the 1st- to the 3rd-order Stokes emission by adjusting the pump power and auto-optimization of the smart filter. Our research uses artificial-intelligence controlled light manipulation in a fiber platform with multi-eigenmodes and nonlinear gain, mapping the spatial control into the spectral domain as well as extending the linear control of light in MMF to active light emission, which is of great significance for applications in optical communication, sensing, and spectroscopy.

## 1. Introduction

The combinations of artificial intelligence and lasers provide powerful ways to form smart light sources with ground-breaking functions. Significantly, the rise of intelligent light control in recent years provides a powerful tool for modulating normal and complex lasing in spatial, spectral, and temporal domains, e.g., studying interactions and dynamics in fiber systems, such as SRS and four-wave mixing [1], temporal-spatial control of pulse [2], and multi-dimensional control of MMF lasers [3].

Fiber lasers with wide wavelength tunability, multi-wavelength, high spectral intensity, and arbitrary spectral shapes are widely needed in diverse areas, such as communication, medicine, optical computing, and imaging [4-7]. Generally, fiber lasers utilize gains from rare-earth-doped fibers, stimulated Brillouin scattering, or stimulated Raman scattering (SRS) and utilize feedback provided by fiber Bragg gratings or fibers loop with filters which jointly determine the spectral characteristics of output. Especially, taking advantages of cascaded SRS and distributed Rayleigh scattering [7-10], half-opened cavity Raman fiber lasers (RFL) can emit at arbitrary wavelength across the fiber transparent window. For example, a high-order RFL has been demonstrated with a continuous wavelength tuning range from 1 μm to 1.9 μm through cascaded Stokes emission pumped by a wavelength-tunable source [11], confirming the full-bandwidth performance of RFL. However, the emission wavelength of these high-order RFLs is mainly controlled by the pump wavelength. Thus, only the single-wavelength with maximum net gain gets emitted, i.e., the spectrum of high-order Stokes lasing cannot be designed freely by the pump. Active spectral shaping that can simultaneously control different orders of Stokes emission is of great importance but is yet to be developed.

A direct way to solve this problem is by introducing a programmable spectral filter into RFLs. There are two methods to realize a programmable spectral filter using spatial light modulator (SLM). The filter realized in method 1 is the combination of SLM and traditional diffraction grating [12]. A grating first disperses the input light to illuminate different spectral components on an SLM, which feedback wanted spectral components by selectively activating the individual spatial elements. Actually, this is a one-dimensional spectral-to-spatial mapping with limited and contradictory spectral resolution and bandwidth that is determined by the pixel size and pixel number of the SLM. Method 2 combines SLM and scattering media [13-15]. Scattering media has a spectrally dependent response with a spectral correlation bandwidth. Beyond the bandwidth, light with different wavelengths passing through the scattering medium will generate uncorrelated speckle patterns. This one-to-one correspondence between the spatial and spectral in scattering media allows spectral control by wavefront shaping of the input light. Compared with traditional gratings, scattering media can realize two-dimensional spectral-to-spatial mapping with high spectral resolution and remarkably extended bandwidth. The drawback of this filter is that its minimum bandwidth is inversely related to the thickness of the scattering media, which will lead to a significant loss when a narrow-band filter is required. Besides, it is not compatible with fiber systems.

Interestingly, multimode fiber (MMF) can perform as a particular scattering medium due to strong intermodal interference, which provides a promising platform for light transportation and manipulation in quasi-turbid media with well light confinement and minimal loss beneficial from its waveguide structure. Various devices have been developed based on random scattering in MMFs, like spectrometers [16,17], endoscopes [18], and optical sensors [19]. Especially, multi-dimensional control of MMF fiber lasers are flourishing [3, 20, 21]. It is no doubt that MMF-base fiber system would have the potential to simultaneously control different orders of Stokes lasing without the problems mentioned above, which have not been studied.

To this end, this work proposed and realized an RFL with reconfigurable and programmable spectra in an ultra-wide bandwidth based on spectral-spatial control of light in multimode fiber (MMF). In the RFL, arbitrary spectral filtering is realized by combining an MMF and an SLM, which can focus light of wanted wavelength(s) into the active part of the laser through wavefront shaping, generating reconfigurable and programmable laser spectra that are controlled by genetic algorithm. By adjusting the pump power and auto-optimization of the filtering effect, smart spectral shaping of the proposed laser from the 1st- to the 3rd- orders of Stokes emission was demonstrated. Differ from existing works, the proposed work uses artificial-intelligence controlled light manipulation in a fiber platform with multi-eigenmodes and nonlinear gain, mapping the spatial control into the spectral domain as well as extending the linear control of light to active light emission, which is of great significance for applications in optical communication, sensing, and spectroscopy.

## 2. Principle and experimental setup

When monochromatic light launches into an MMF, a unique speckle pattern will generate at its output due to multi-mode interference. The speckle pattern is related to the input wavelength, which is the basis of MMF-based filtering. This section will discuss numerically and experimentally the spectral correlation bandwidth (SCB) of MMF, that mainly depends on structural parameters of MMF, i.e., length, numerical aperture, and core diameter [16]. The whole paper uses the step-index-MMF with a core/cladding diameters of 105/125 μm and a numerical aperture of 0.22, so we mainly focus on the effect of length on the SCB. The electric field at the output end of MMF read:

$$E(r,\phi,\lambda,z) = \sum_{n=1}^{N} A_n e_n(r,\phi,\lambda) e^{-i(\beta_n(\lambda)z - \omega t + \varphi_n)} \tag{1}$$

where $e_n$ and $\beta_n$ are the spatial profile and propagation constant of the $n$th guided mode of the MMF, calculated by solving Maxwell's equations [22]. $A_n$ and $\varphi_n$ are the amplitude and initial phase of the $n$th guided mode of the MMF, respectively. It is assumed that all the guided modes in the MMF were excited equally, with initial phases randomly distributed between 0 and $2\pi$. Input light of different wavelengths will result in different speckle patterns at the output end of the MMF due to the superposition of guided modes with wavelength-sensitive accumulated phases. White circles in Fig. 1(a) show the process of speckle decorrelation in a 1.5 m-long MMF when the input wavelength changes by 0.1 nm. The spectral correlation functions $C(\Delta\lambda)$ versus wavelength change, $\Delta\lambda$, for different lengths of MMF are calculated numerically in Fig. 1(b), from which the SCB, $\Delta\lambda_C$ (i.e., full-width half maximum of the curves), are plot in Fig. 1(c). It reflects that the SCB decreases with the length increasing of MMF.

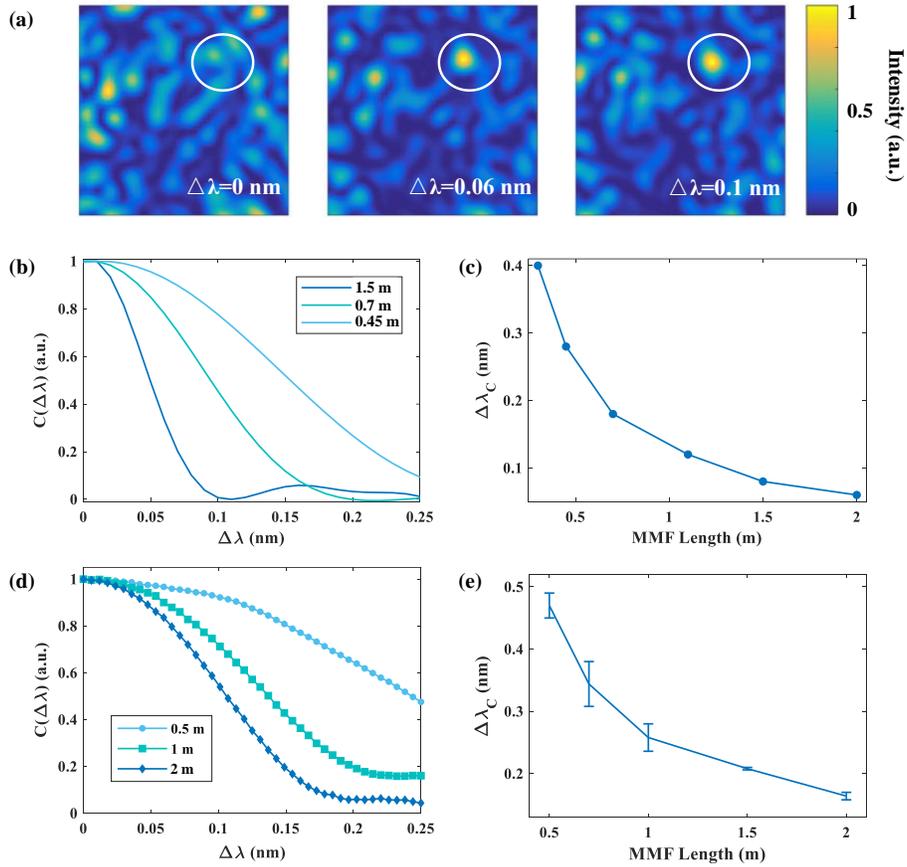

Fig. 1. Wavelength-dependent speckle patterns of the MMF. (a) Normalized speckle patterns at different input wavelengths for MMF length of 1.5 m. (b) Simulated and (d) measured spectral correlation functions for MMFs with different lengths. (c) Simulated and (e) measured spectral correlation bandwidth for MMFs with different lengths.

Similarly, $C(\Delta\lambda)$ and $\Delta\lambda_C$ are measured experimentally using a tunable laser and a near-infrared camera. Based on the speckle patterns obtained at wavelengths from 1549.5 nm to 1550.5 nm, the results are given in Figs. 2(d) and 2(e), respectively, showing good accordance with the numerical results.

Obviously, wavelengths of the light beyond the SCB can be mapped one-to-one to speckle patterns, which can be used to develop an optical filter through spatial light modulation. The bandwidth precision of the filter will be determined by the length of MMF, i.e., the value of SCB.

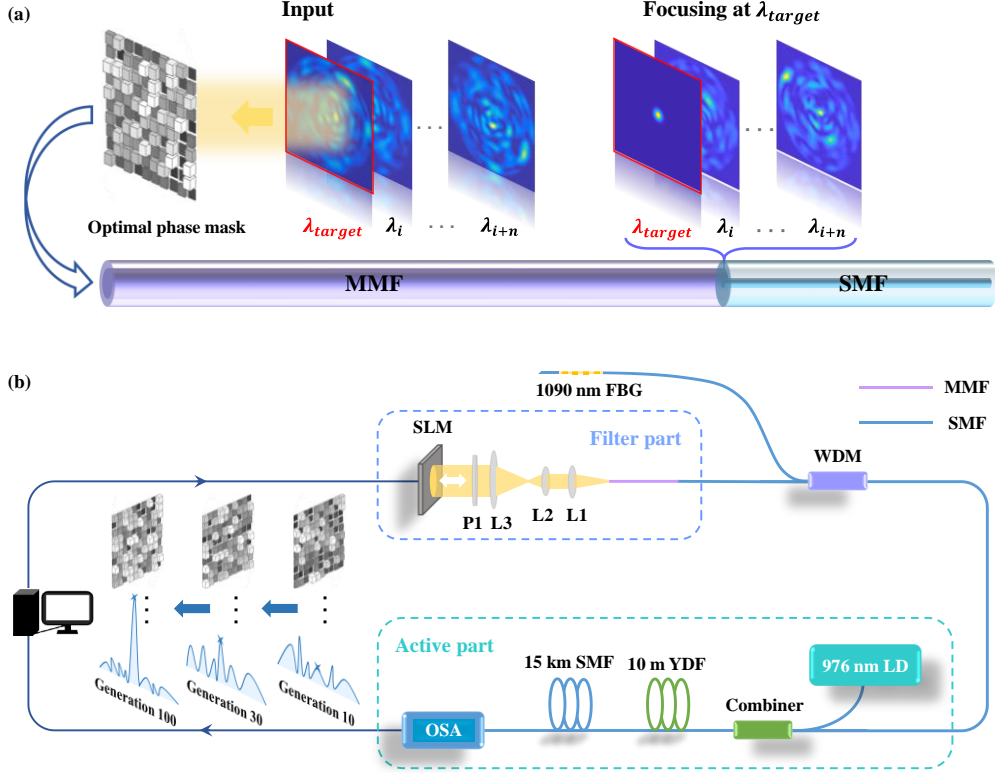

Fig. 2. Working mechanism and experimental setup. (a) Working mechanism and (b) schematic diagram of the spectral reconfigurable Raman fiber laser. FBG, fiber Bragg grating; WDM, wavelength division multiplexer; LD, laser diode; YDF, Ytterbium-doped fiber; SLM, spatial light modulator; P1, polarizer; L1, L2, L3, lens; SMF, single-mode fiber; MMF, multimode fiber; OSA, optical spectrum analyzer. The inset shows optimization process.

The working mechanism and experimental setup of the spectral reconfigurable and programmable RRFL are shown in Fig. 2. The fiber laser includes a filter part and an active part. The filter part works as a powerful wavelength programmable point reflector that feedbacks the light of Raman scattering as seed light to control the laser emission. Light from the active part is first coupled into a segment of MMF, collimated by lens 1 (L1) and expanded by a telescope (L2, L3, 10×). Then it is polarized and illuminated on a phase-only SLM (Meadowlark Optics, P1920-500-1200-HDMI) to shape the light wavefront that feedbacks into the MMF. The MMF is spliced with the single-mode fiber (SMF) of the active part. Due to modal field mismatch between the MMF and SMF, the spectral profile of the feedback into the active part is determined by the part of speckle pattern at the MMF-SMF splicing interface that is coupled into the SMF (i.e., modal overlapping). Thus, the filter can work by modulating wavelength-correlated speckle patterns of the MMF through the SLM-controlled wavefront shaping, which can focus light of desired wavelength(s) from the MMF into the SMF and can be controlled intelligently by a genetic algorithm. Light from the filter part is then injected into the active part and determines the lasing spectra.

The active part, together with the filter, form an RFL with a forward-pumping and half-opened structure. In the active part, an Ytterbium-doped random fiber laser (YRFL) is firstly achieved, where the gain is provided by a 976 nm laser diode (LD) pump, and the feedback is provided by a 1090 nm fiber Bragg grating and Rayleigh scattering in the SMF. Then cascaded Raman random lasing can be realized based on the pump of the realized 1090 nm YRFL and wavelength programming of the filter.

Intelligent control of the laser output is implemented through iterative optimization of the genetic algorithm. The laser output is monitored by an optical spectrum analyzer (OSA), and its difference from the target spectrum is made into the objective function of the algorithm, which is used to update the phase pattern of the SLM. By doing this iteratively, the SLM can focus the light with the target wavelength(s) into the active part of the laser while blocking unwanted light feedback as much as possible (seeing the inserted speckle patterns at the MMF-SMF splicing interface in Fig. 2(a)). Because the feedback light acts as seed light that finally determines the output wavelength, only a small power (e.g., larger than spontaneous emission) of feedback back is enough to control the output, as also verified in previous works [9, 23]. Once the SLM pattern is found for laser output, it can be recorded and reloaded to generate the specific laser output as long as the MMF keeps undisturbed. By doing this, the output spectrum can switch as fast as the frame rate of the SLM (e.g., 30 Hz in our case). Moreover, during the running of the laser, the optimization can be repeated to eliminate the slow degeneration of the state of MMF.

## 3. Results and Discussion

We first investigated the characteristics of the reflective filter. Fig. 3(b) shows an example of single wavelength filtering at 1555.5 nm, wherein the filtered spectra before and after intelligent optimization of the SLM's pattern are both given. The spectrum before the optimization is noisy and disordered due to the randomness of the wavelength-dependent speckle pattern. The spectrum, after optimization, establishes a peak at the target wavelength of 1555.5 nm with a peak-to-ground ratio of 13 dB. As studied in the previous part, the length of the MMF determines the SCB, which can be used to set the bandwidth precision of filtering. The insert of Fig. 3(a) shows that the measured bandwidths precision of filtering decreases with the length increasing of the MMF, which matches well with the conclusion of Fig. 1. We adopt the genetic algorithm to optimize the filter and take the intensity value at 1555.5nm as the objective function. The fitness value grows fast with iteration increasing until the convergence state is reached at about the 800th iteration, seeing Fig. 3(b).

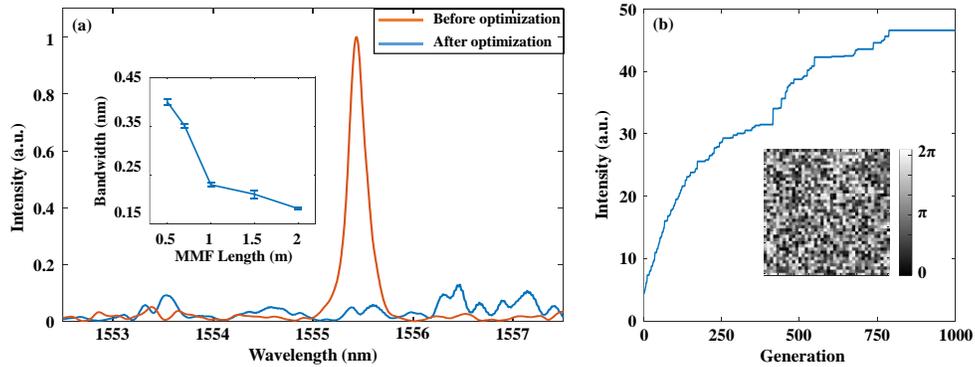

Fig. 3. Performance of the filter part. (a) Output spectrum before and after optimization and the optimized wavelength is 1555.5 nm. The inset shows the bandwidth of the filter as a function of the length of the MMF. (b) Optimization process corresponding to (a). The inset shows the optimal phase pattern.

Then the spectral programmable cascaded RFL is achieved by using the filter part together with the active part. The wide reflection band of the filter part and the cascaded SRS effect in the active part both enable a high degree of freedom for laser control in the spectral domain. We first achieved a continuously tunable single-wavelength RFL at the 1st-order Stokes band. Typical spectra of lasing are given in Fig. 4(a). The MMF-length-determined bandwidth of lasing is about 0.4 nm, and the signal-to-noise ratio (SNR) of the spectrum is greater than 30 dB, indicating high spectral purity. The output power of the 1st-order single-wavelength RFL at 1145 nm versus the LD pump power is shown in Fig. 4(b), reflecting that the lasing threshold is 3.8 W and the conversion efficiency is about 11%. Actually, the quantum efficiency of the laser can be increased remarkably at the expense of increasing the lasing threshold if a shorter length of SMF in the active part is used.

Moreover, multi-wavelength lasing with arbitrary wavelength separation and envelope can be realized by setting the target spectra shape as the objective function of the genetic algorithm. Two examples are shown in Figs. 4(c) and 4(d), which are dual-wavelength and five-wavelength lasing with designed wavelength separations and power distribution. This performance verifies the flexibility of laser manipulation in the spectral domain and is quite different from filters based on Fabry-Perot cavities or SMF-MMF-SMF structures wherein the spectral shape cannot be designed freely.

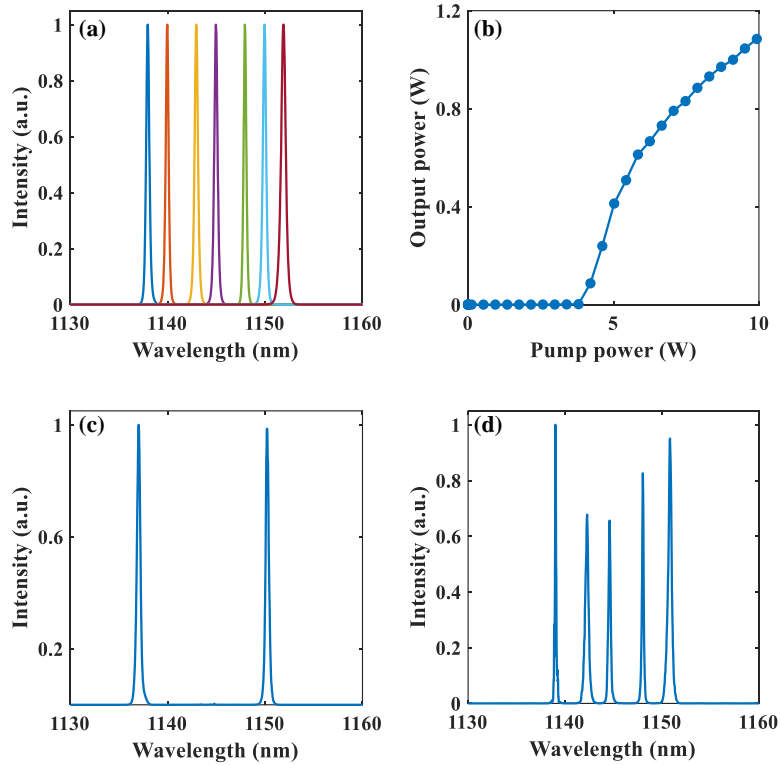

Fig. 4 Output characteristics of 1st-order random Raman fiber laser. (a) Tunable single-wavelength output. (b) The output power of Raman lasing versus LD pump power. (c) Dual-wavelength and (d) Five-wavelength output.

In addition, existing filters cannot cover an extensive wavelength band while keeping a relatively high spectral precision. Thus, simultaneously tuning in wavelength of different orders of Raman lasing can only be realized by varying the pump wavelength with limited control in spectral shape and bandwidth unless multiple filters of different bands are used together. Our method can overcome this limitation. As a further functional demonstration of the proposed laser, spectral programmable and reconfigurable cascaded RFL is achieved in Fig. 5. To control lasing of different orders of Strokes emission, we optimize the lasing spectrum in several

stages—according to the cascaded order of Raman lasing. For the 1st-order Raman lasing, the phase pattern of the SLM is optimized to minimize the difference between the output spectrum and the target spectrum. The final phase pattern of the SLM in the previous stage was taken as the starting point for optimizing the higher-order Raman lasing. Figs. 5(a) and 5(b) show the continuously tunable single-wavelength lasing from the 1st- to 3rd-order emission. Here, we can tune the lasing wavelength of higher-order emission while keeping the wavelength of previous order unchanged or changed.

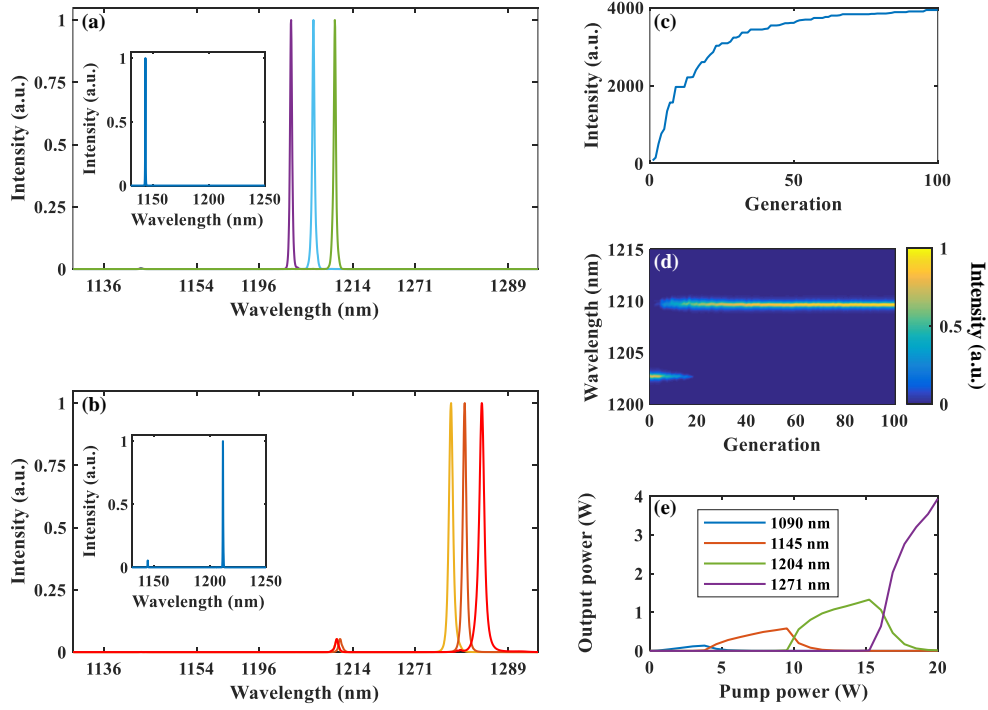

Fig. 5 Reconfigurable and programmable Raman lasing. (a) Wavelength tunability of the 2nd-order RFL. The inset is the corresponding 1st-order RFL. (b) Wavelength tunability of the 3rd-order RFL. The inset shows the corresponding lower-order RFL. (c) Intensity evolution of the 2nd-order RFL using genetic algorithm. (d) Evolution of wavelength switching from 1203 nm to 1210 nm of the 2nd-order RFL. (e) Output power of the cascaded RFL versus LD pump power.

In Fig. 5(a), the 1st-order Raman lasing is targeted at 1143 nm (the inset) at the pump power of 9 W. Then, the pump power increases to 13.5 W, and the phase pattern of the SLM is set to different optimized patterns to emit 2nd-order Raman lasing of different wavelengths. Besides, the running Raman laser can also dynamically switch the output wavelength through new target optimization. As shown in Figs. 2(c) and 2(d), the wavelength of 2nd-order Raman lasing switches from 1203 to 1210 nm after a small round of iterative optimization. Similarly, the 3rd-order Raman lasing with variable wavelength can be realized through 3-stage optimizations, e.g., in Fig. 3(b), the first two stages optimize the 1st- and 2nd-order lasing at 1143 nm and 1211 nm, respectively, and the third stage optimizes the 3rd-order lasing at the target wavelengths. Fig. 5(e) shows that the output power transfers between lower-order and higher-order Raman lasing, taking single-wavelength lasing at 1145, 1204, and 1271 nm as an example. The energy purity of transitions is larger than 99.9%. Because the optimization is continuously order-by-order, lasing efficiency increases with the order of emission and the bandwidth of lasing can always keep around 0.4 nm. This reflects that our optimization is a nonlinear control of lasing other than a linear control of a passive filter. All of these reflect that the spectrum of our proposed laser can be controlled simultaneously and independently in an ultra-wide wavelength band with a programmable spectral shape and almost un-broadened laser bandwidth.

Because the proposed laser's output is goal-oriented, the iteration can dynamically optimize the output according the state change of MMF, e.g., degeneration of the transmission matrix of the MMF due to the instability of the external environment (mechanical disturbance, temperature change, etc.). This can always keep the laser working stably. As shown in Fig. 6, the lasing spectrum is optimized in real-time by monitoring the intensity value of the target wavelength and setting the degradation threshold. Once the value falls below the threshold, the genetic iteration is rerun to ensure that the filter is always working optimally. Taking the single-wavelength lasing of the 1st-order RFL as an example at a fixed pump power of 9W, we test the time stability of the target wavelength at 1143nm, including both spectrum and output power. We first run the genetic algorithm to achieve single-wavelength output, and in the next 9 hours, use the spectrometer and power meter to record the output. When spectral degradation is detected, the spectrum can be recovered only after running a few iterations, demonstrating the high stability and utility of our laser.

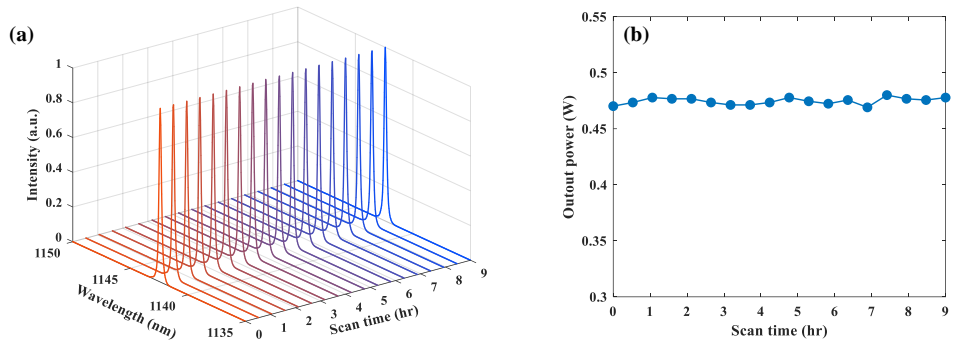

Fig. 6 Stability of the laser. (a) Temporal stability of the modulated lasing spectrum. (b) The time stability of output power corresponding to (a).

## 4. Conclusion

In conclusion, based on wavelength-sensitive spatial focusing of light in a multimode fiber by wavefront shaping, a smart spectral programmable Raman fiber laser is proposed with ultra-wide tunable band and arbitrary spectral shape. The laser includes a reflective filter part based on intelligent spectral-spatial control of light in MMF and an active part that provides cascaded Raman gain and distributed Rayleigh feedback, which can emit reconfigurable and programmable spectral of Raman lasing from the 1st- to the 3rd-orders. The laser has a wide operation band and arbitrary shaping of output spectral while keeping a relatively narrow lasing bandwidth, which has not been reported before to the best of our knowledge. This work would brave an intelligent way to actively control light in nonlinear systems and provide powerfully controllable light sources for optical information applications.

**Funding.** This work is supported by the National Natural Science Foundation of China under Grants 11974071 and 61635005